# Honeycomb Phononic Networks with Closed Mechanical Subsystems

Xinzhu Li, Mark C. Kuzyk, and Hailin Wang

Department of Physics, University of Oregon, Eugene, OR 97403, USA

Abstract

We report the design of a diamond-based honeycomb phononic network, in which a mechanical resonator couples to three distinct phononic crystal waveguides. This two-dimensional (2D) phononic network extends an earlier study on one-dimensional (1D) phononic networks with closed mechanical subsystems. With a special design for the phononic band structures of the waveguides, any two neighboring resonators in the 2D network and the waveguide between them can form a closed mechanical subsystem, which enables nearest neighbor coupling and at the same time circumvents the scaling problems inherent in typical large mechanical systems. In addition, the 2D network can be attached to a square phononic crystal lattice and be protected by the large band gap of the phononic crystal shield. Honeycomb phononic networks of spin qubits with nearest neighbor coupling can serve as an experimental platform for quantum computing and especially topological quantum error corrections.

## I. INTRODUCTION

Mechanical waves cannot propagate in vacuum and are thus immune to scattering losses into vacuum. This, along with the relatively slow phase velocity and relatively long wavelength, gives mechanical networks a distinct advantage over optical networks for on-chip quantum networks[1-5]. Mechanical or phononic networks have been pursued as a platform for developing quantum computers that can take advantage of robust spin qubits such as color centers in diamond[1,4,5]. Coherent interactions between acoustic waves and a variety of qubit systems, such as semiconductor quantum dots[6-10], superconducting circuits[11-14], nitrogen vacancy centers in diamond [15-28], and spin defect centers in SiC[29], have been actively explored.

Large mechanical networks, including one-dimensional chains of trapped ions, however, have inherent and well-known scaling problems[30]. Nearest neighbor coupling of a large number of mechanical resonators can lead to the formation of spectrally dense mechanical modes. The crosstalk between these modes prevents the quantum control of individual mechanical modes. In addition, the speed of gate operations in these networks scales inversely with the size of the mechanical system. These scaling problems, which are unavoidable in a large mechanical network, can be circumvented in a network architecture, which breaks the large mechanical network into small and closed mechanical subsystems [5]. This architecture has been implemented theoretically with the use of alternating phononic crystal waveguides in a one-dimensional (1D) phononic network, in which a mechanical resonator couples to two distinct phononic crystal waveguides[5]. While the phononic waveguides enable the coupling between neighboring resonators[31], the specially designed phononic bandgaps in the alternating waveguides prevent the propagation of the relevant mechanical vibration to adjacent waveguides.

In this 1D phononic network, any two neighboring resonators and the waveguide between them can form a closed mechanical subsystem. Quantum state transfers or two-qubit quantum operations between spin qubits in neighboring resonators can take place in the closed mechanical subsystem via phonon-assisted or sideband spin transitions driven by either optical or microwave fields, as shown in an earlier study[5]. This phononic network of spin qubits feature effectively nearest neighbor couplings mediated by a phononic waveguide, but without the scaling problems in typical large mechanical networks.

In this paper, we extend the earlier work on 1D phononic networks to two-dimensional (2D) phononic networks by designing a honeycomb phononic network that features closed mechanical subsystems. The building block of our 2D phononic network is a single mechanical resonator coupling to three distinct phononic crystal waveguides. We have developed a design principle for the phononic band structures of the waveguides such that in this 2D network, any two neighboring resonators and the waveguide between them can form a closed mechanical subsystem. To isolate and protect the relevant mechanical modes of the 2D network from the surrounding environment, we have also proposed an approach for attaching the honeycomb network to a phononic crystal shield. Our numerical design is based on the use of diamond, for which both photonic and optomechanical crystals have been successfully fabricated[32-35]. The design can be easily extended to other material systems.

Honeycomb phononic networks of spin qubits with nearest neighbor coupling can be used as an experimental platform for quantum computing, including topological quantum computing. While topological quantum error correction schemes such as the surface code have focused on the use of 2D square quantum networks with nearest neighbor coupling[36-39], the honeycomb network is of special importance for exploring and understanding topological quantum computing and especially quantum error corrections[40]. Furthermore, this type of networks can serve as an experimental platform for studies of topological quantum excitations such as anyons[41].

**II. HONEYCOMB PHONONIC NETWORKS**

For the 1D phononic network of alternating waveguides developed in the earlier study[5], waveguide and resonator modes in the non-overlapping spectral regions of the phononic band gaps of the alternating waveguides are used. In this case, the phononic band gaps prevent the propagation of the relevant mechanical vibrations in the network and confine the vibrations in a local mechanical subsystem. To extend this concept to higher dimensions, in which a mechanical resonator couples directly to more than two distinct phononic waveguides, we require that a mechanical mode employed in the network can propagate in only one of the distinct waveguides. This mode thus needs to be in the phononic band gaps of other waveguides.

For communications among qubits within the same mechanical resonator, a mechanical mode that is in the band gap of all waveguides is preferred, as discussed in the earlier study[5]. This means an additional requirement that the phononic band gaps of the waveguides involved should have an overlapping or common spectral region, in which no mechanical modes can propagate.

For a 2D network, a given mechanical resonator can couple to three or four phononic waveguides. A square, or more generally, rectangular lattice can be formed with the use of four waveguides. Similarly, a honeycomb lattice can be formed with the use of three waveguides. In this paper, we focus on the honeycomb network, since it is easier to satisfy the requirements discussed above with three waveguides in a honeycomb lattice than with four waveguides in a rectangular lattice.

## A. Building blocks of the honeycomb lattice

A natural choice for mechanical resonators in a honeycomb phononic network is a thin triangular plate with equal side length $s$. For additional flexibility in tuning mechanical frequencies of the resonator, we cut off the three corners, which have a side length of $s'$, from the triangular plate, as illustrated in Fig. 1a. The phononic network employs the symmetric compressional modes of the thin plate, i.e. modes that are symmetric with respect to the median plane of the plate. For a fixed $s$, we can adjust the frequencies of these mechanical modes by varying $s'$.

As shown in Fig. 1a, three phononic waveguides, $A$, $B$, and $C$, are connected to the mechanical resonator along the three symmetry axes of the resonator. These waveguides feature a periodic array of elliptical holes and are effectively 1D phononic crystals[31]. As illustrated in Fig. 1b, the width of the waveguide is $w$ and the distance between neighboring holes is $d$. The semi-major axis and the semi-minor axis of the elliptical hole are $a$ and $b$, respectively. Figure 1c depicts a honeycomb lattice composed of these elementary units, for which waveguides $A$ and $B$ have the same length, with $L_A = L_B$, while waveguides $C$ have a different length, $L_c$. For the numerical calculation of a diamond-based phononic network presented in this paper, we take the thickness of the 2D structure to be 0.3 μm.

## B. Phononic band structures

For a honeycomb phononic network with closed mechanical subsystems, we need three waveguide modes propagating in waveguides $A$, $B$, and $C$, respectively. Each mode is forbidden in the other two waveguides. The three waveguides also need to have a shared or common band gap region, in which no mechanical modes can propagate in any of the waveguides. These requirements are impossible to satisfy, if we consider only a single band gap for each waveguide. However, the above requirements can be satisfied, if within the spectral region of interest, one of the waveguides feature two phononic band gaps, while the other two still have only a single phononic band gap.

Figure 2 shows the phononic band structures of waveguides *A*, *B,* and *C* for the symmetric compression modes, calculated with the COMSOL Multiphysics software package and with $d$ = 6, 4, 7.6 µm for waveguides *A*, *B* and C, respectively. Within the spectral region of interest, waveguides *A* and *B* feature single phononic band gaps, while waveguide *C* features two phononic band gaps. For these calculations, we have taken (*w*, *a*, *b*) to be (3, 1.1, 0.3), (3, 1.1, 0.3), (2, 0.8, 0.76) µm for waveguides *A*, *B* and C, respectively. Figure 2 highlights four special spectral regions. Region I is in the band gap of waveguides *B* and *C*, but not *A*. Region II is in the band gap of waveguides *A* and *C*, but not *B*. Region III is in the band gap of waveguides *A* and *B*, but not *C*. In this case, waveguide modes in Regions I, II, and III can only propagate in waveguides *A*, *B* and C, respectively. Region IV is a common band gap for all three waveguides. Resonator modes in this region can be used for coupling among qubits within a given resonator[5].

The band diagram and especially the selection of the four special spectral regions shown in Fig. 2 depend on the geometric parameters of the waveguide. There are inevitable deviations or fluctuations from the design parameters in the fabrication of the waveguides. Figure 3 shows, as an example, how the two band gaps in waveguide *C* change as we vary the four geometric parameters, *d*, *w*, *a*, and *b*, by up to ± 100 nm. As shown in the figure, the lower band gap is more stable against parameter variations than the upper band gap and the strongest dependence occurs for the geometric parameters of the elliptical holes, including the semi-major axis, *a*, and the semi-minor axis, *b*.

For the four special spectral regions shown in Fig. 2, the stability of the lower band gap of waveguide C is of more importance than that of the upper band gap. The lower band gap affects the boundaries for regions II, III, and IV, with region III being the narrowest among the four spectral regions. In addition, there are no spectral separations between regions II, III, and IV. In comparison, region I is spectrally separated from the other three regions and is the widest among the four spectral regions. In this regard, our design of the four spectral regions for the desired waveguide operations should be able to tolerate considerable variations in the geometric parameters of the waveguides.

## C. Closed mechanical subsystems

The nearest neighbor coupling in our phononic network takes place via closed mechanical subsystems, which consist of any two neighboring resonators and the waveguide between them. The waveguide modes in this closed mechanical subsystem are standing mechanical waves. In the limit of relatively short waveguides, we can approximate a given waveguide as a single-mode waveguide. Because of the relatively small resonator-waveguide coupling rate, quantum state transfer between spins in neighboring resonators requires that the resonator mode is nearly resonant with the corresponding waveguide mode. Specifically, we need three resonator modes, with frequency $\omega_a$, $\omega_b$, and $\omega_c$, to be nearly resonant with the corresponding waveguide modes in waveguide $A$, $B$, and $C$, respectively.

The frequency of a resonator mode depends on the side length $s$ and $s'$ as well as the spatial displacement pattern. The frequency of a waveguide mode depends on the waveguide length and other geometric parameters including $d$, $w$, $a$, $b$. We can tune the frequency of a waveguide mode without affecting the phononic band structure by varying the length of the waveguide. For the honeycomb lattice discussed in this paper, we have chosen $L_A=L_B$ (see Fig. 1c). This means variations of the waveguide lengths alone cannot tune independently the frequencies of all three relevant waveguide modes. Additional tuning parameters, such as $s$, $s'$, $d$, $w$, $a$, $b$, are thus needed in order to satisfy the resonance condition for all three sets of resonator and waveguide modes.

Figures 4a, 4b, and 4c show the displacement patterns of three resonator modes that can couple to one of the three distinct phononic waveguides and are in the band gap of the other two waveguides. The frequencies of the three modes are respectively $\omega_a/2\pi=1.7339$ GHz, $\omega_b/2\pi=0.9634$ GHz, and $\omega_c/2\pi=1.3388$ GHz. With $L_A=L_B=86.3$ μm and $L_C=91.2$ μm, these resonator modes are resonant with the corresponding waveguide modes featuring $\omega_A/2\pi=1.7339$ GHz, $\omega_B/2\pi=0.9634$ GHz, and $\omega_C/2\pi=1.3388$ GHz (the difference between the resonator and the corresponding waveguide frequency is of order a few kHz). The geometric parameters of the waveguides are the same as those used in Fig. 2. The mode spacing is 31, 37, 28 MHz for waveguides $A$, $B$, and $C$, respectively. The displacement pattern of a mechanical mode, whose frequency, $\omega_d/2\pi=1.1691$ GHz, is in the band gap of all three waveguides (region IV in Fig. 2), is displayed in Fig. 4d.

For a closed mechanical subsystem containing two neighboring resonators and the waveguide between them, the resonator-waveguide coupling leads to the formation of three normal modes. In the single-mode waveguide limit, we can use numerically-determined normal mode frequencies to deduce the effective detuning, $\Delta$, between the resonator and waveguide modes and the resonator-waveguide coupling rate $g$, according to [5]

$$\Delta = \omega_+ + \omega_- - 2\omega_0, \tag{1a}$$

$$g = \sqrt{\frac{(\omega_+ - \omega_-)^2 - \Delta^2}{8}}, \tag{1b}$$

where $\omega_+$, $\omega_-$, and $\omega_0$ are the frequencies of the three normal modes. Note that with $\Delta=0$, the normal mode with frequency $\omega_0$ is a dark mode, for which there is no waveguide mode component. The corresponding resonator-waveguide coupling rates are $g_A/2\pi=4.3$ MHz, $g_B/2\pi=19$ MHz, and $g_C/2\pi=5$ MHz.

The resonator-waveguide coupling rate for waveguide $B$ is much greater than those for the other two waveguides because the displacement pattern shown in Fig. 4b exhibits the strongest displacement in the contact area between the resonator and the waveguide. In this case, $g_B/2\pi$ derived from Eq. 1 is about 1/2 of the mode spacing for waveguide $B$, indicating that the single-mode approximation is no longer adequate. Note that the coupling rate can be reduced by tailoring the end of the waveguide.

Figure 4e plots the displacement patterns of the three normal modes of the closed mechanical subsystem associated with waveguide $C$. The normal mode in the middle is approximately a dark mode. Figure 4f shows the displacement pattern of this normal mode in a local region of the phononic network. As expected, the mechanical excitation is confined in the closed mechanical subsystem containing the waveguide $C$ and the two resonators coupled directly to the waveguide. Note that the frequency of the dark mode shown in Fig. 4e differs slightly (by about 1 MHz) from the corresponding waveguide and resonator mode frequency calculated separately for the individual structures. In addition, the dark mode is not completely dark, even with the zero-detuning deduced from Eq. 1. These behaviors are likely the result of residual coupling to nearby waveguide modes.

The relevant mechanical frequencies of a phononic network depend sensitively on the dimensions and geometric parameters of both the resonators and the waveguides. The achievement of the resonance between the resonator and the corresponding waveguide modes thus presents a formidable technical challenge for the experimental realization of the phononic network.

In this regard, for the various schemes on the quantum state transfer or two-qubit quantum operations in a closed mechanical subsystem[5], the schemes that can tolerate considerable detuning between the resonator and the corresponding waveguide modes are preferred. It should also be noted that for phononic networks with closed mechanical subsystems, the resonance conditions are only required for local mechanical subsystems. Our architecture is robust against global or long-range variations in the resonator and waveguide frequencies so long as the local resonance conditions are satisfied.

**III. PHONONIC CRYSTAL SHIELD**

While it was straightforward to use a square phononic crystal lattice as a phononic shield for 1D phononic networks as well as single mechanical resonators[5,42-45], additional modifications are needed in order to embed a honeycomb phononic network in a phononic crystal lattice because of the special geometry and symmetry of the honeycomb network. To provide the needed isolation and protection for the honeycomb phononic network, we have used a square phononic crystal lattice as a phononic shield and have made minor modifications in the square lattice, while retaining the protection provided by the relevant phononic band gaps.

The unit cell of a square phononic crystal lattice, with a period of 3.8 μm, is shown in Fig. 5a. The phononic band structure of the symmetric compression modes of the square lattice features a large band spanning from 0.85 to 2.35 GHz (see Fig. 5b), which can provide excellent protection for all the relevant mechanical modes in our phononic network. Figure 6 shows schematically the attachment of a honeycomb phononic network to a square phononic crystal lattice. The phononic structure highlighted by the dotted line box in Fig. 6 shows that phononic crystal waveguide $C$ with a period of 7.6 μm is embedded directly in the square phononic crystal lattice with bridges or linkers of suitable length. In this case, the phononic band gap of the waveguide also provide additional protection for the relevant mechanical modes in the network.

The phononic structure highlighted by the dashed line box in Fig. 6 consists of a section of the square phononic crystal lattice attached to phononic crystal waveguide $C$ near the bottom of the structure. Note that adjacent sections of square phononic crystal lattices can be connected together at the boundary with bridges of suitable length. Given the relatively large size of these sections, the section boundaries will have negligible effects on the phononic waveguides.

## IV. CONCLUSIONS

In summary, we have extended the earlier theoretical work of 1D phononic networks with closed mechanical subsystems to 2D phononic networks by using a honeycomb lattice and by designing special band structures for the phononic crystal waveguides. For the 2D honeycomb network, a closed mechanical subsystem can be formed from any two neighboring resonators and the waveguide between them. These closed subsystems enable the nearest neighbor coupling in the 2D network and overcome the scaling problems inherent in large mechanical systems. Experimental realization of this type of 2D phononic networks of spin qubits can open up a new platform for pursuing fault-tolerant quantum computers and for exploring topological quantum error corrections.

Finally, we note that 2D square or rectangular phononic networks with closed mechanical subsystems, which can allow the implementation of the surface code, can be realized with each mechanical resonator coupling to four distinct phononic crystal waveguides. The required band structures for the phononic crystal waveguides and the resonance condition for the relevant resonator and waveguide modes will be more difficult to achieve than those for the honeycomb networks.

## ACKNOWLEDGEMENTS

This work is supported by AFOSR and by NSF under grants No. 1606227 and No. 1641084.

# APPENDIX: CALCULATION OF PHONONIC BAND STRUCTURES AND MECHANICAL MODES

We have carried out finite element numerical calculations with COMSOL Multiphysics to analyze the phononic band structures of 1D and 2D phononic crystal structures and to determine the frequencies and field patterns of mechanical normal modes in a mechanical resonator. For variations or sweeps of geometric parameters, we have combined the use of COMSOL Multiphysics and MATLAB. The diamond material parameters used in the numerical calculations include Young's modulus $E = 1050$ GPa, Poisson ratio $\nu=0.2$, and the material density $\rho = 3539$ kg/m$^3$.

For the calculation of mechanical modes in an elastic material, we treat these modes as a continuum field with a time-dependent displacement, $\mathbf{Q}(\mathbf{r}, t)$, which satisfies the wave equation,

$$\rho \partial_t^2 \mathbf{Q} = (\lambda + \mu)\nabla(\nabla \cdot \mathbf{Q}) + \mu \nabla^2 \mathbf{Q}, \qquad (2)$$

where the Lamé constants are given by

$$\lambda = \frac{\nu E}{(1+\nu)(1-2\nu)}, \quad \mu = \frac{E}{2(1+\nu)}. \qquad (3)$$

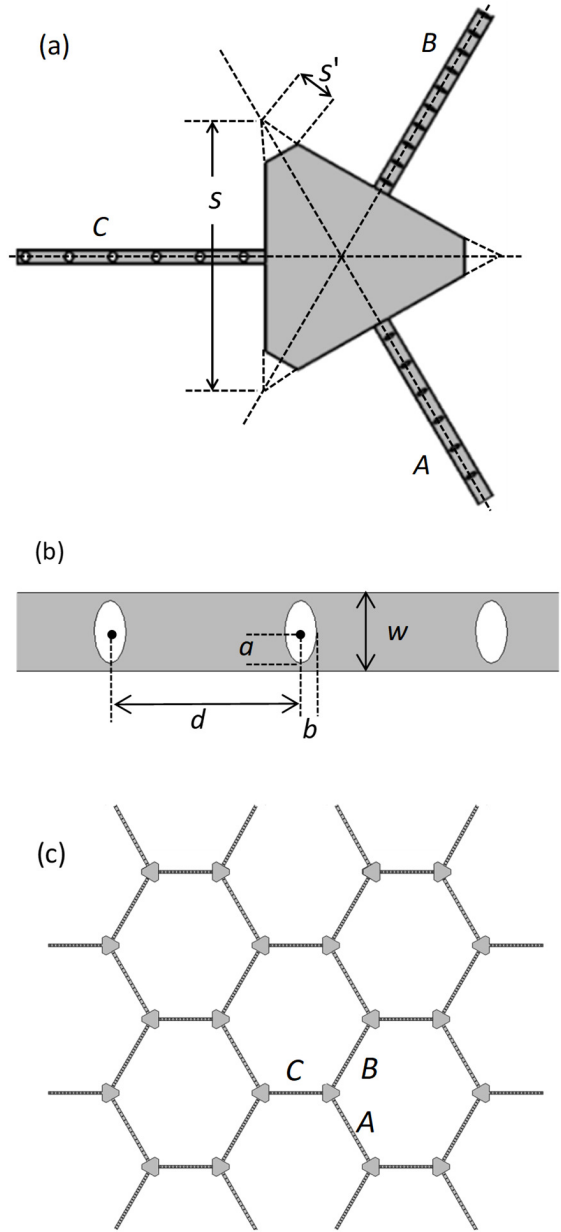

**Figure 1.** (a) The building block of the honeycomb phononic network, for which an equilateral triangular mechanical resonator, with the three corners cut off, couples to phononic crystal waveguides, A, B, and C, along three symmetry axes of the resonator. (b) The geometry of the phononic crystal waveguide. For the numerical calculations, we have taken ($d$, $w$, $a$, $b$) to be (6, 3, 1.1, 0.3) μm, (4, 3, 1.1, 0.3) μm, and (7.6, 2.0, 0.8, 0.76) μm for waveguides A, B and C, respectively. (c) A honeycomb lattice, for which waveguides A and B have the same length, while waveguide C has a different length.

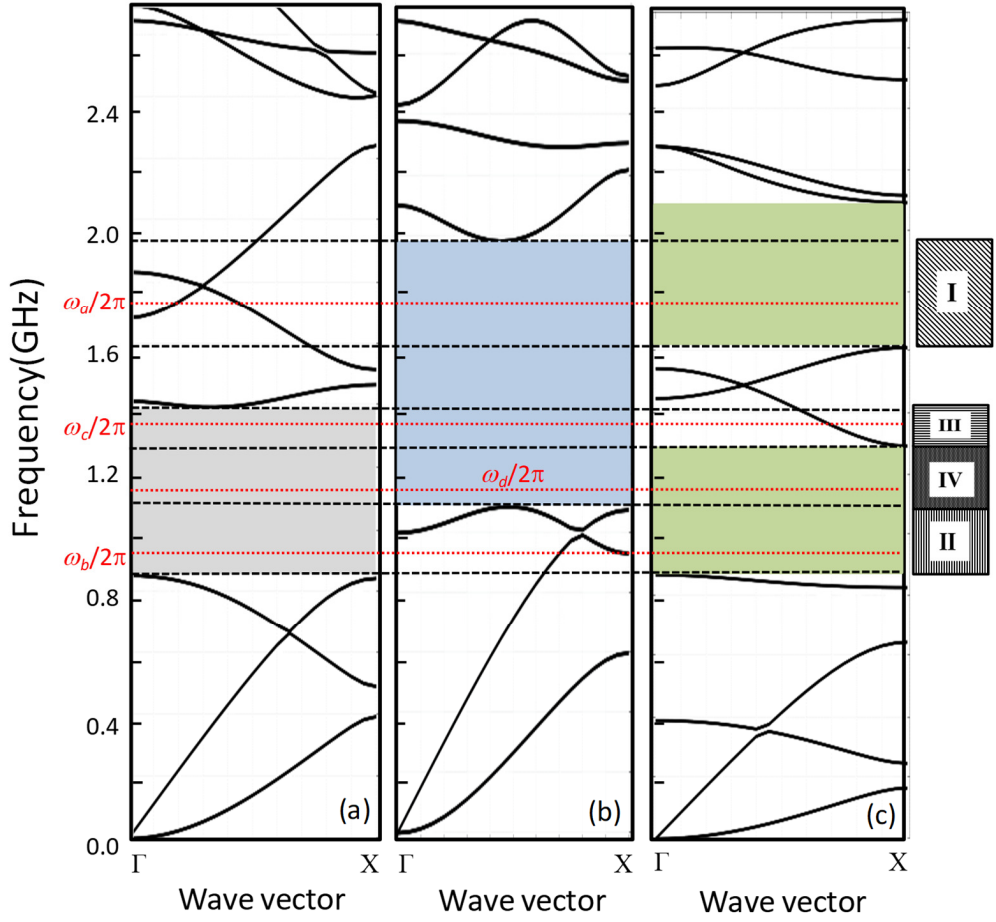

**Figure 2.** (a), (b), (c) The phononic band structures of the symmetric compression modes for waveguides *A*, *B*, and *C*, respectively. The shaded areas highlight the band gaps. The dashed lines mark four important regions. Region I is in the band gaps of waveguides *B* and *C*, but not *A*. Region II is in the band gaps of waveguides *A* and *C*, but not *B*. Region III is in the band gaps of waveguides *A* and *B*, but not *C*. Waveguide modes in Regions I, II, and III can only propagate in waveguides *A*, *B* and *C*, respectively. Region IV is a common band gap for all three waveguides. The dotted lines indicate the frequencies of the four resonator modes shown in Fig. 4.

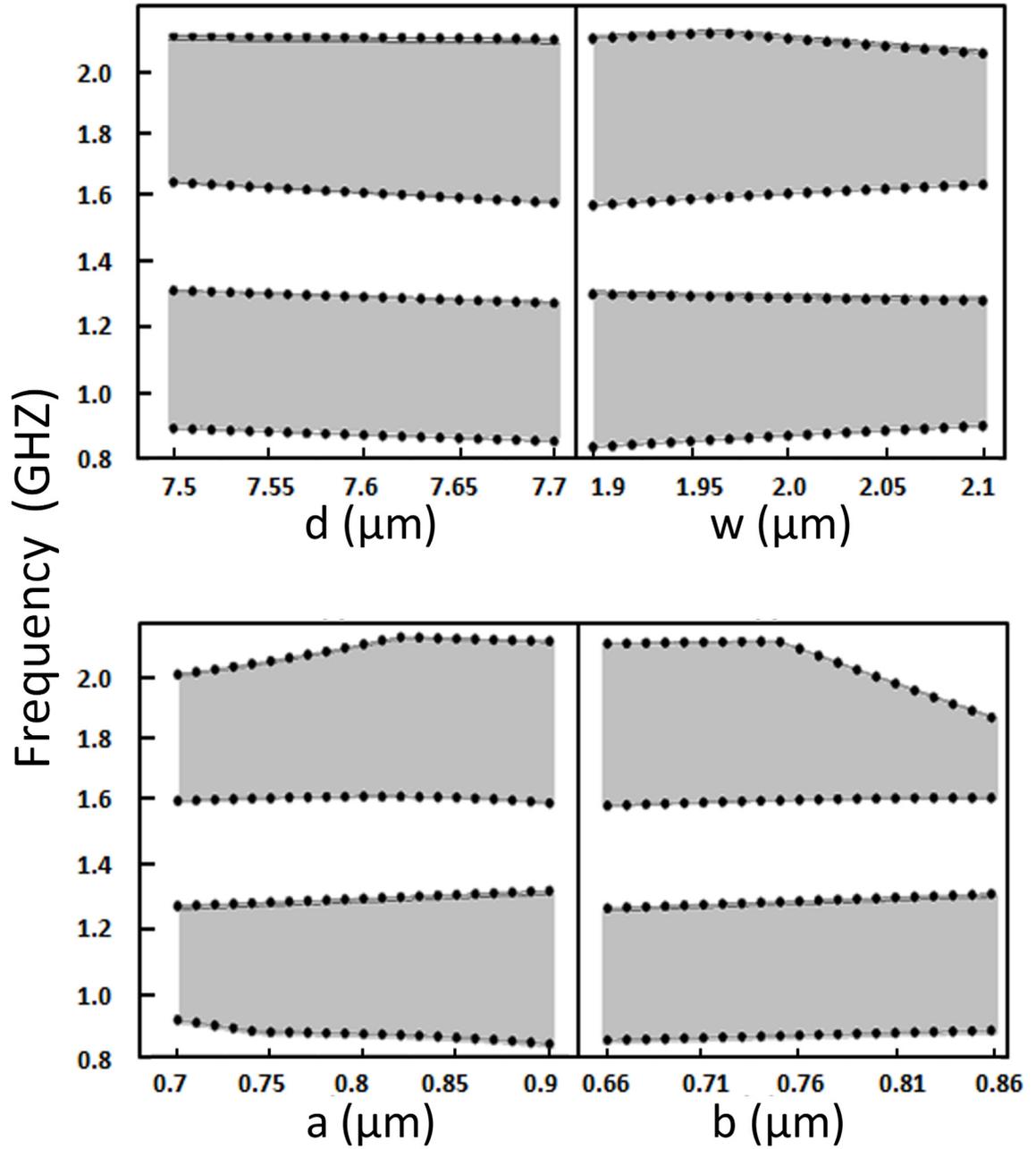

**Figure 3.** The two band gaps (highlighted by the shaded areas) of waveguide $C$ as a function of four waveguide parameters, $d$, $w$, $a$, and $b$. These parameters are varied by up to $\pm 100$ nm around the desired value, $(d, w, b, a) = (7.6, 2.0, 0.8, 0.76)$ μm.

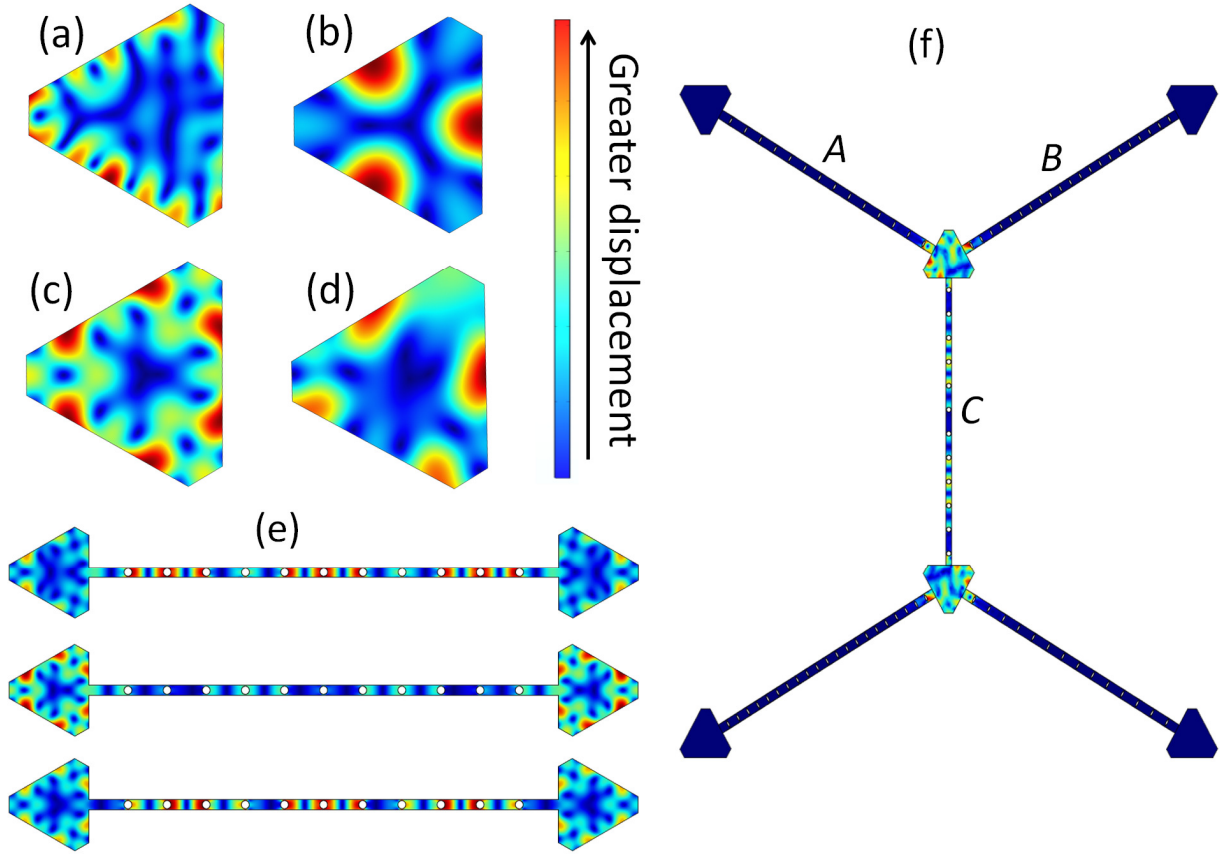

**Figure 4.** (a), (b), (c), (d) The displacement patterns of mechanical resonator modes with $\omega_a/2\pi$=1.7339 GHz, $\omega_b/2\pi$=0.9634 GHz, $\omega_c/2\pi$=1.3388GHz, $\omega_d/2\pi$=1.1691 GHz, respectively, and with $s$=21 μm and $s'$=3.15 μm. The modes with frequency, $\omega_a$, $\omega_b$, $\omega_c$, are resonant with the corresponding waveguide modes in waveguides $A$, $B$, $C$, respectively. The mode with frequency $\omega_d$ is in the gap of all three waveguides. (e) The displacement patterns of the three normal modes associated with waveguide $C$ with $L_C$=91.2μm. The frequencies of the three modes are 1.3326, 1.3397, 1.3468 GHz from top to bottom. (f) The displacement pattern of the normal mode with frequency $\omega/2\pi$=1.3397 GHz in a local region of the phononic network. This mode, which can propagate in waveguide $C$, is forbidden in waveguides $A$ and $B$, leading to a closed mechanical subsystem containing waveguide $C$ and the two resonators coupled directly to the waveguide.

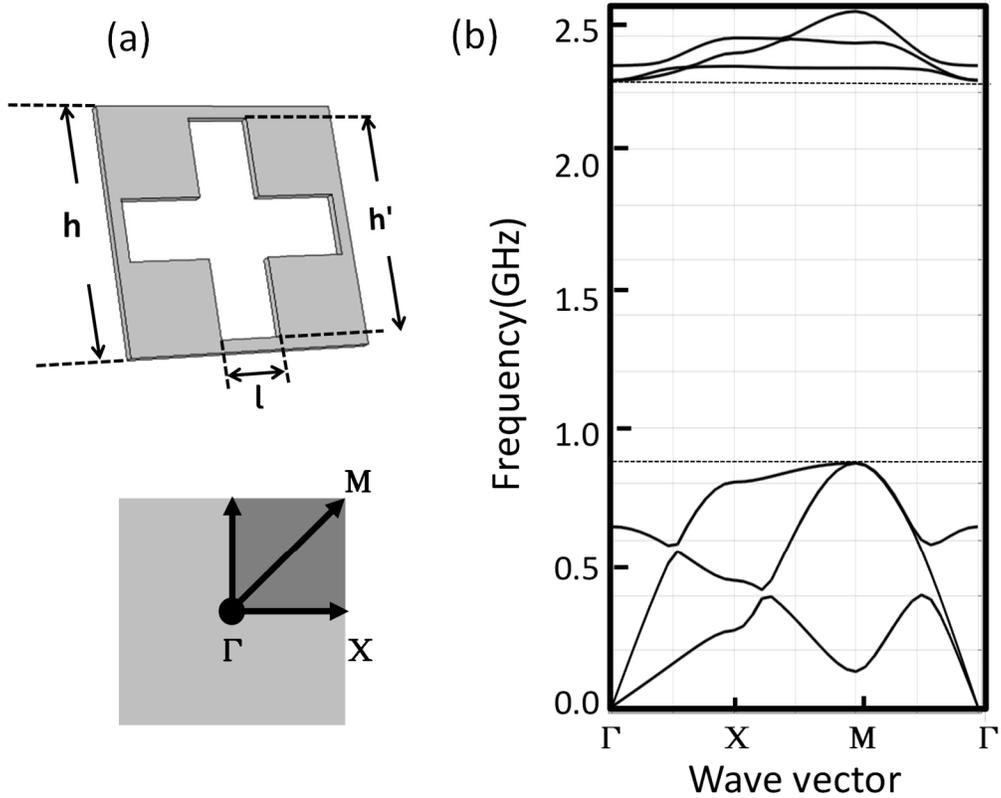

**Figure 5.** (a) The unit cell of a square phononic crystal lattice with a period of 3.8 μm, along with the reciprocal space of a square lattice. The lattice parameters are $h$=3.8μm, $h'$=3.5μm, and $l$ =1μm. (b) The phononic band structure of the symmetric compression modes of the square lattice, featuring a large gap spanning from 0.85 to 2.35 GHz.

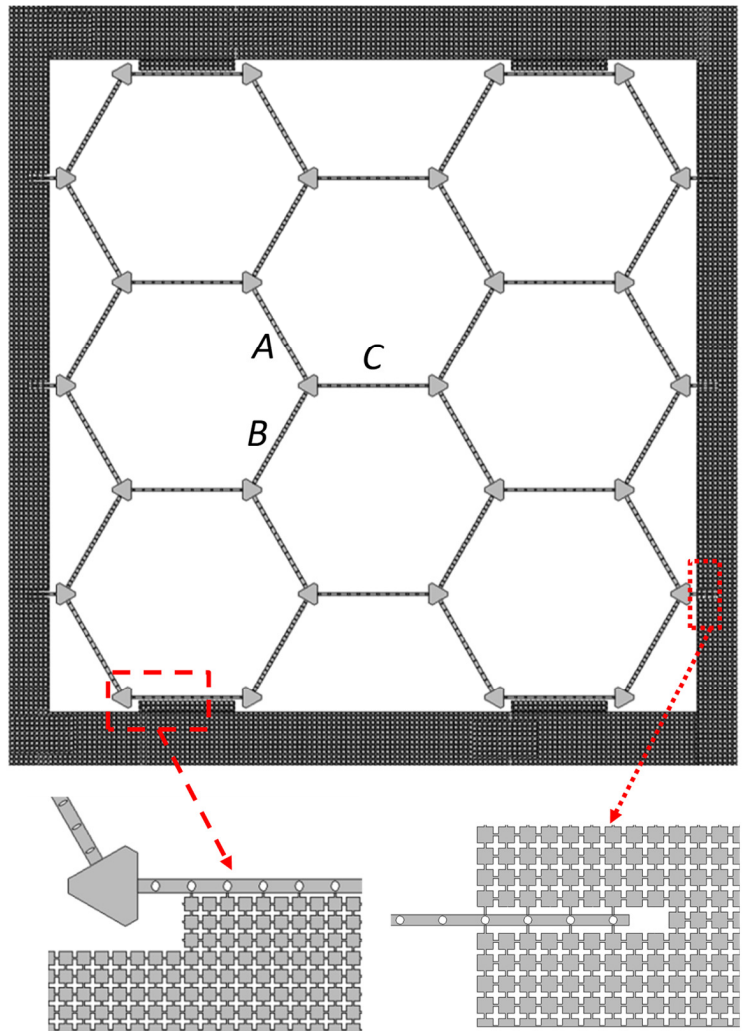

**Figure 6.** Schematic of the honeycomb phononic network attached to the square phononic crystal shield shown in Fig 5. The expanded drawings show in detail the attachment of the waveguide *C* with a period of 7.6 μm to the square phononic crystal lattice.